\documentclass[a4paper,11pt]{article}

\usepackage{pos}
\usepackage{booktabs}
\usepackage[noabbrev]{cleveref}
\usepackage[acronym,shortcuts]{glossaries-extra}
\setabbreviationstyle[acronym]{long-short}
\glsdisablehyper
\usepackage[italic]{hepnames}
\usepackage{lineno}
\usepackage{physics}
\usepackage{siunitx}
\usepackage{subcaption}
\usepackage{slashed}
\usepackage{bm}

\newacronym{gpd}{GPD}{generalized parton distribution}

\AtBeginDocument{\RenewCommandCopy\qty\SI}
\DeclareUnicodeCharacter{0302}{^}

\title{Lattice QCD extraction of the $\Peta_{c}$-meson $t$-dependent parton distribution function}
\ShortTitle{LQCD extraction of the $\Petac$-meson tPDF}

\author[a]{B. Blossier}
\author[b]{C. Mezrag}
\author*[b]{J.M. Morgado Ch\'{a}vez}
\author[a]{T. San Jos\'{e}}

\affiliation[a]{Laboratoire de Physique des 2 Infinis Irène Joliot-Curie, CNRS/IN2P3,\\
Université Paris-Saclay, 91405 Orsay Cedex, France}

\affiliation[b]{D\'{e}partement de Physique Nucl\'{e}aire, Irfu/CEA-Saclay,\\
91191 Gif-sur-Yvette Cedex, France}

\abstract{The formalism of short-distance factorization, conveyed through the pseudo-distribution approach, connects space-like and light-cone correlators and thus allows for the extraction in lattice QCD of a number of parton distributions. We compute the $t$-dependent parton distribution function of valence quarks in a $\Peta_{c}$-meson. After reviewing the main ideas behind the pseudo-distribution formalism, and relying on the analytic structure of Ioffe-time distributions, we come up with a proposal for a model-independent extraction of $t$-dependent parton distribution functions. We present results for the $\Peta_{c}$-meson Ioffe time valence $t$-dependent parton distribution function at a renormalization scale of $3~\textrm{GeV}$.}

\FullConference{31st International Workshop on Deep Inelastic Scattering (DIS2024)\\
8–12 April 2024\\
Grenoble, France\\}

\begin{document}
\maketitle

\section{Introduction}

Pseudoscalar mesons are bound states of quarks and gluons as well as Nambu-Goldstone modes associated to the dynamical breakdown of chiral symmetry. This characteristic makes them an ideal laboratory to study fundamental phenomena such as the emergence of hadron mass. In this sense, exploring the structure of pseudoscalar mesons becomes an interesting research subject, motivating the work presented in this proceeding. 

We focus on the case of the $\Peta_{c}$-meson, which is a $J^{P}=0^{-}$ state arising from a pair $c\bar{c}$ of valence quarks and thus constitutes a good framework for the study of quark-mass effects in the emergence of hadronic mater. In the pursuit of this objective we seek for a window to the internal structure of the $\Peta_{c}$-meson. Here the different parton distributions are of special relevance. In this work we target the leading-twist $t$-dependent parton distribution function (tPDF) of quarks within an $\Peta_{c}$ state  \cite{Diehl:2003ny,Belitsky:2005qn}:
\begin{equation}\label{eq:tPDF-def}
q_{\Peta_{c}}(x,t,\mu^{2})=\frac{1}{2}\int_{-\infty}^{\infty}\frac{d\nu}{2\pi\nu}e^{-i\nu x}\langle\Peta_{c}(p+\Delta_{\perp})|\bar{c}(0)\slashed{z}[0;z]c(z)|\Peta_{c}(p-\Delta_{\perp})\rangle,
\end{equation}
where $\nu=-p\cdot z$, often dubbed Ioffe time; $z^{\mu}=(z^{3},0,0,z^{3})$ and $2\Delta_{\perp}^{\mu}=(0,\Delta^{1},\Delta^{2},0)$. The straight Wilson line $[0;z]$ renders the quantity invariant under gauge transformations.

As it is apparent from Eq.~\eqref{eq:tPDF-def}, these objects arise from the generalized parton distributions (GPDs) when the momentum transfer between hadron states is restricted to occur on the plane transverse to the direction joining quark fields. Thus, they are much simpler than GPDs, simplifying their calculation; but they still draw an exhaustive picture of hadron's inside, providing access to standard PDFs and electromagnetic form factors or allowing to perform hadron tomography \cite{Burkardt:2000za}.

The main goal of this proceeding is to obtain the tPDF of quarks within an $\Peta_{c}$-meson Eq.~\eqref{eq:tPDF-def}, laying the groundwork for a study of quark-mass effects in the emergence of hadron structure. Being non-perturbative objects, we work in the framework of lattice QCD (LQCD), aiming at the \textit{ab-initio} evaluation of tPDFs. We start summarizing the formalism allowing to extract tPDFs from LQCD calculations (Sec.~\ref{sec:Formalism}). Finally, in Sec.~\ref{sec:LQCDsetup}, we describe our numerical calculation and its results.

\section{Method}\label{sec:Formalism}

As all parton distributions, tPDFs are defined from matrix elements of non-local operators involving a light-like separation between fields, c.f. Eq.~\eqref{eq:tPDF-def}. For a long time, this feature constituted the main challenge for their direct evaluation in Euclidean setups such as LQCD. Many efforts were devoted to circumvent this issue, crystallizing in a number of successful strategies (see \textit{e.g.} \cite{Cichy:2021ewm}). We choose the pseudo-distribution approach \cite{Radyushkin:2017cyf} where a generalization of the matrix element in Eq.~\eqref{eq:tPDF-def} to spacelike $z$ is considered
\begin{equation}\label{eq:matele}
\begin{array}{rcl}
\displaystyle 2\nu M_{\Peta_{c}}(p,\Delta_{\perp},z) &\displaystyle= &\displaystyle\langle\Peta_{c}(p_{\textrm{out}})|\bar{c}(0)\slashed{z}[0;z]c(z)|\Peta_{c}(p_{\textrm{in}})\rangle\\
\\
&\displaystyle = &\displaystyle 2\nu\mathcal{M}_{\Peta_{c}}(\nu,t,z^{2})+(\Delta_{\perp}\cdot z)\mathcal{G}_{\Peta_{c}}(\nu,t,z^{2})+z^{2}\mathcal{Z}_{\Peta_{c}}(\nu,t,z^{2}),
\end{array}
\end{equation}
where we label $p_{\textrm{out}}=p+\Delta_{\perp}$ and $p_{\textrm{in}}=p-\Delta_{\perp}$.

This matrix element shows UV divergences originated by the space-like Wilson line which can be absorbed by a multiplicative renormalization factor depending solely on the quark-antiquark separation \cite{Ishikawa:2017faj}. We shall then consider its renormalized counterpart, $M(p,\left.\Delta_{\perp},z)\right|_{\textrm{R}}=Z^{-1}(z)M(p,\Delta_{\perp},z)$, and explore its small-$z^{2}$ behavior. To this end we employ a light-cone operator product expansion:
\begin{equation}\label{eq:OPE}
\left.2\nu M_{\Peta_{c}}(p,\Delta_{\perp},z)\right|_{\textrm{R}}\xrightarrow{z^{2}\to 0}\sum_{i}\sum_{n=0}^{\infty}C_{n}^{(i),\overline{\textrm{MS}}}(z^{2},\mu^{2})\langle\Peta_{c}(p_{\textrm{out}})|\mathcal{O}_{(i)}^{\{z\mu_{1}\cdots\mu_{n}\}}(0)|\Peta_{c}(p_{\textrm{in}})\rangle_{\overline{\textrm{MS}}}\prod_{k=1}^{n}z_{\mu_{k}}+\textrm{ h.t.}
\end{equation}
where $i$ labels the different twist-two operators $\mathcal{O}^{\{z\mu_{1}\cdots\mu_{n}\}}\equiv\mathcal{O}^{\{\mu\mu_{1}\cdots\mu_{n}\}}z_{\mu}$ \cite{Belitsky:2005qn} and ``$\textrm{h.t.}$'' denotes higher-twist contributions. The subscript $\overline{\textrm{MS}}$ indicates that the divergences developed by the Wilson coefficients $C_{n}$ as $z$ approaches the light-front are handled in the corresponding scheme.

We now consider the Lorentz covariant decomposition of the matrix elements of twist-two operators. For definiteness, we focus on operators of quark-type\footnote{This can be achieved using combinations with definite quantum numbers, $C$-odd, for the matrix element. As a result, the \textit{valence} sector of the hadrons is accessed. However, trying to simplify the discussion, we do not go through this point.} only, writing
\begin{equation}\label{eq:OPEcoefs}
\left.2\nu M_{\Peta_{c}}(p,\Delta_{\perp},z)\right|_{\textrm{R}}\xrightarrow{z^{2}\to 0} 2\nu\sum_{n=0}^{\infty}C^{\overline{\textrm{MS}}}_{n}(z^{2},\mu^{2})(1+z^{2}f_{n}(\nu,t,z^{2}))a_{n}^{\overline{\textrm{MS}}}(t,\mu^{2})(2\nu)^{n}+\mathcal{O}(z^{2})+\textrm{ h.t.}
\end{equation}
where $a_{n}$ represent Lorentz invariant amplitudes arising in the decomposition of the matrix elements of twist-two operators; $f_{n}$ are given functions of a purely kinematic origin and the extra $\mathcal{O}(z^{2})$ terms weight coefficients other than $a_{n}$. Owing to the previous expansion, the light-front contribution given by the renormalized matrix element is found to read
\begin{equation}\label{eq:matching}
\begin{array}{rcl}
\displaystyle \widetilde{q}_{\Peta_{c}}(\nu,t,z^{2})\equiv\left.\mathcal{M}_{\Peta_{c}}(\nu,t,z^{2})\right|_{\textrm{R}} & \displaystyle = & \displaystyle \sum_{n=0}^{\infty}C^{\overline{\textrm{MS}}}_{n}(z^{2},\mu^{2})a_{n}^{\overline{\textrm{MS}}}(t,\mu^{2})(2\nu)^{n}+\cdots\\
\\
&\displaystyle = &\displaystyle \int_{0}^{1}dw~\mathfrak{C}(w,z^{2},\mu^{2})q_{\eta_{c}}(w\nu,t,\mu^{2})+\cdots
\end{array}
\end{equation}
where the ellipses labels terms proportional to $z^{2}$ and $q_{\Peta_{c}}(\nu,t,\mu^{2})$ is the Ioffe time tPDF \cite{Radyushkin:2017cyf,Braun:1994jq}
\begin{equation}
q_{\Peta_{c}}(\nu,t,\mu^{2})=\int_{-1}^{1}dx e^{i\nu x}q_{\Peta_{c}}(x,t,\mu^{2}).
\end{equation}

The left-hand side of the matching relation Eq.~\eqref{eq:matching}, the pseudo-distribution \cite{Radyushkin:2017cyf}, can be realized as the extension off the light-front for standard Ioffe time distributions and, importantly, can be computed for space-like separations. The matching kernel $\mathfrak{C}$ can be computed in perturbation theory \cite{Radyushkin:2017cyf}. Exploiting the matching relation Eq.~\eqref{eq:matching} in combination with lattice data for the pseudo-tPDF allows therefore to extract the desired tPDF.

There is, however, one further difficulty: the matching kernel acts on light-cone distributions and ``pushes'' them off the light-front while they are the pseudo-distributions which might be obtained from LQCD. Attempting at the inversion of such relation would introduce further complications at the time that it would require the introduction of \textit{Ans\"atze} allowing to incorporate a discrete set of data points into an integral relation. In the spirit of \cite{Bhattacharya:2023ays}, we propose to work directly in Ioffe time space and express both, pseudo-distributions and light-cone distributions, as power series in the Ioffe time variable:
\begin{equation}\label{eq:PowerSeries}
\widetilde{q}_{\Peta_{c}}(\nu,t,z^{2})=\sum_{k=0}^{\infty}\widetilde{a}_{k}(t,z^{2})\nu^{n},\qquad q_{\Peta_{c}}(\nu,t,\mu^{2})=\sum_{k=0}^{\infty}a_{k}(t,\mu^{2})\nu^{n}.
\end{equation}

In fact, because pseudo- and light-cone distributions are defined as Fourier transforms of distributions with compact support $x\in[-1,1]$ \cite{Radyushkin:1983wh}, they can be shown to be analytic functions in the Ioffe time variable\footnote{To the best of our knowledge there is only evidence, \textit{e.g.} \cite{Izubuchi:2018srq}, of parton distributions to be tempered distributions.} (see \textit{e.g.} Ch.~$7$ in \cite{Strichartz:1994mt}). In this sense, the parametrizations suggested in Eq.~\eqref{eq:PowerSeries} are strictly model independent and indicate that $x$-space distributions must be represented as power series in the Dirac delta distribution and its derivatives \cite{Zhang:2024djl}. Furthermore, when plugged into the matching relation Eq.~\eqref{eq:matching}, the cumbersome convolution reduces to a product
\begin{equation}
\widetilde{a}_{k}(t,z^{2})=c_{k}(z^{2},\mu^{2})a_{k}(t,\mu^{2}),\qquad c_{k}(z^{2},\mu^{2})=\int_{0}^{1}dw w^{k}\mathfrak{C}(w,z^{2},\mu^{2}),
\end{equation}
which allows to connect pseudo-distribution coefficients to light-cone ones, as required in practice. We can now define a strategy allowing a model-independent extraction of tPDFs --and, with minor modifications any other parton distribution-- from LQCD:
\begin{itemize}
\item Fit pseudo-distribution data to a power-series
\begin{equation}\label{eq:FitAnsatz}
\widetilde{q}^{\textrm{data}}_{\Peta_{c}}(\nu,t,z^{2})=\sum_{k=0}^{N}(1+z^{2}\widetilde{B}_{k})\widetilde{A}_{k}(t,z^{2})\nu^{n}.
\end{equation}
where we also take into account target-mass corrections through an additional fitting coefficient, $\widetilde{B}$, which effectively takes into account the corresponding behavior, Eq.~\eqref{eq:OPEcoefs}.

\item Match the extracted coefficients, $\left.\widetilde{a}_{k}(t,z^{2})=\widetilde{A}_{k}(t,z^{2})\right|_{\textrm{Cont.}}$ to the light-front as $\displaystyle a_{k}(t,\mu^{2})=\widetilde{a}_{k}(t,z^{2})/c_{k}(z^{2},\mu^{2})$.
\item Reconstruct the light-cone $\overline{\textrm{MS}}$ Ioffe time distribution.
\end{itemize}

\section{Lattice calculation}\label{sec:LQCDsetup}

With a roadmap for the extraction of tPDFs from a LQCD calculation, we are left with the task of generating the necessary data for the pseudo-tPDF. To this end we dispose a numerical setup relying on a set of $N_{f}=2$ ensembles generated by the \texttt{CLS} effort \cite{Fritzsch:2012wq} (see Tab.~\ref{tab:cls-ensembles}). We arrange a kinematic configuration with hadron momenta $p_{\textrm{in}}^{\mu}=(E_{\textrm{in}},\bm{p}_{\perp},p^{3})$, $p_{\textrm{out}}^{\mu}=(E_{\textrm{in}},-\bm{p}_{\perp},p^{3})$, and $z^{\mu}=(0,0,0,z^{3})$. Thereupon, using the sequential propagator technique, we evaluate
\begin{equation}\label{eq:3pt}
C_{3}^{ss'}(\vec{p}_{\textrm{in}},\vec{p}_{\textrm{out}},t_{\textrm{src}},\tau)=\sum_{\vec{x},\vec{z}}e^{-i\vec{p}_{\textrm{in}}\cdot\vec{x}}e^{-i\vec{\Delta}_{\perp}\cdot\vec{z}}\langle\widehat{\Peta}_{c}^{s}(\vec{x},t_{\textrm{src}})\bar{c}(\vec{z}_{\Delta\vec{z}},\tau)\gamma^{0}[\vec{z}_{\Delta\vec{z}};\vec{z}]_{\tau}c(\vec{z},\tau)\widehat{\Peta}_{c}^{s'}(\vec{0},0)\rangle
\end{equation}
where $\vec{z}_{\Delta\vec{z}}=\vec{z}+\Delta\vec{z}$, and $s$, $s'$ label different smearing levels. We focus on the gamma structure $\gamma^{0}$ which, in the kinematic setup employed and, according to the Lorentz covariant decomposition in Eq.~\eqref{eq:matele}, yields access to the desired amplitude, $\mathcal{M}$.

\begin{table}[t]
\centering
\begin{tabular}{lccc cc ccc}
\toprule
  id &
  $\beta$ &
  $a~[\unit{\femto\metre}]$ &
  $L/a$ &
  $am_{\Ppi}$ &
  $m_{\Ppi}~[\unit{\mega\eV}]$ &
  $m_{\Ppi}L$ &
  $\kappa_{\Pcharm}$ &
  $\kappa_{l}$ \\
\midrule
  A5  & 5.2 & 0.0755(9)(7) & $32$ & \num{0.1265(8)} & 331 & 4.0 & 0.12531 & 0.13594 \\
\midrule
  E5  & 5.3 & 0.0658(7)(7) & $32$ & \num{0.1458(3)} & 437 & 4.7 & 0.12724 & 0.13625 \\
  F7  &     &              & $48$ & \num{0.0885(3)} & 265 & 4.3 & 0.12713 & 0.13638 \\
\midrule
  N6  & 5.5 & 0.0486(4)(5) & $48$ & \num{0.0838(2)} & 340 & 4.0 & 0.13026 & 0.13667 \\
\bottomrule
\end{tabular}

\caption{Set of \texttt{CLS} ensembles used in this proceeding. From left to right: ensemble label, bare strong coupling, lattice spacing \cite{Fritzsch:2012wq}, spatial extent of the lattice ($T=2L$), approximate value of the pion mass \cite{DellaMorte:2017dyu}, the proxy of finite-volume effects $m_{\Ppi}L$, the value of $\kappa_{\Pcharm}$ \cite{Balasubramamian:2019wgx} and $\kappa_{u}=\kappa_{d}=\kappa_{l}$.}
    \label{tab:cls-ensembles}
\end{table}

For this calculation we use quark all-to-all propagators with wall sources diluted in spin. We use one interpolator for the $\Peta_{c}$ state, $\widehat{\Peta}_{c}=\bar{c}\gamma_{5}c$, and four different APE-blocking Gaussian smearings. The solution of the corresponding generalized eigenvalue problem (GEVP) is used to project the correlation function to a realistic ground-state. Furthermore, we use twisted boundary conditions and we do not consider quark-disconnected diagrams.

Using this setup, data for the necessary three-point functions, Eq.~\eqref{eq:3pt}, are generated on spacetime separations $|\Delta\vec{z}|\in[0,9]$ in units of the lattice spacing. The desired amplitude is then extracted from \textit{plateau} fits to the ratios
\begin{equation}
\frac{C_{3}^{\textrm{P}}(\vec{p}_{\textrm{in}},\vec{p}_{\textrm{out}},t_{\textrm{src}},\tau)}{\sqrt{C_{2}^{\textrm{P}}(\vec{p}_{\textrm{out}},t_{\textrm{src}})C_{2}^{\textrm{P}}(\vec{p}_{\textrm{in}},t_{\textrm{src}})}}\sqrt{\frac{C_{2}^{\textrm{P}}(\vec{p}_{\textrm{in}},t_{\textrm{src}}-\tau)C_{2}^{\textrm{P}}(\vec{p}_{\textrm{out}},\tau)}{C_{2}^{\textrm{P}}(\vec{p}_{\textrm{out}},t_{\textrm{src}}-\tau)C_{2}^{\textrm{P}}(\vec{p}_{\textrm{in}},\tau)}}\xrightarrow{(t_{\textrm{src}}-\tau)>>0}\frac{2\nu\mathcal{M}_{\Peta_{c}}(\nu,t,z^{2})}{4\sqrt{E(\vec{p}_{\textrm{in}})E(\vec{p}_{\textrm{out}})}}
\end{equation}
where the superscript $\textrm{P}$ indicate correlation functions projected according to the solution of a GEVP. The optimal fit range is identified using the Akaike information criterion and model averaging \cite{Jay:2020jkz}. Finally, following the suggestion of \cite{Radyushkin:2017cyf}, the renormalization program is carried out in the form
\begin{equation}
\widetilde{q}_{\Peta_{c}}(\nu,t,z^{2})=\frac{\mathcal{M}_{\Peta_{c}}(\nu,t,z)}{\mathcal{M}_{\Peta_{c}}(0,0,z)}\frac{\mathcal{M}_{\Peta_{c}}(0,0,0)}{\mathcal{M}_{\Peta_{c}}(\nu,t,0)}
\end{equation}
which, in addition, imposes the expected normalization for the pseudo-tPDF. 

\begin{figure}[t]
\centering
\hspace*{-1.8cm}\begin{subfigure}{0.45\textwidth}
\includegraphics[width=1.35\textwidth]{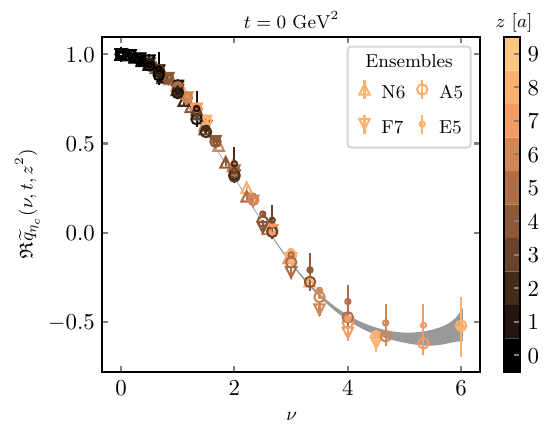}
\end{subfigure}
\hfill
\begin{subfigure}{0.45\textwidth}
\hspace*{-1cm}\includegraphics[width=1.3\textwidth]{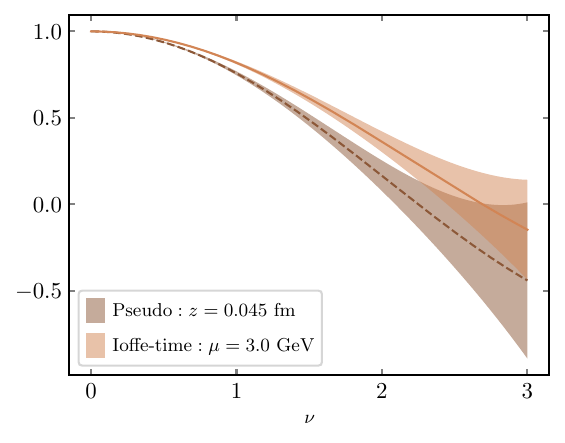}
\end{subfigure}
\caption{\textsc{Left panel}: LQCD data for the valence contribution to the pseudo-tPDF (real part of the pseudo-tPDF) at $t=0~\textrm{GeV}^{2}$ together with their best fit. \textsc{Right panel}: Comparison between the continuum extrapolation of the LQCD valence pseudo-PDF, gray band, and its light-cone limit, red band.}\label{fig:results}
\end{figure}

As a result of this procedure, the pseudo-tPDF data are obtained. For illustration, results for the real part of the pseudo-tPDF in the limit of no momentum transfer and all considered ensembles are shown in Fig.~\ref{fig:results}, left panel. Notably, the signal-to-noise ratio shown by the data remains under control over the entire range of Ioffe times and, as expected, all data-points tend to a universal line.

With reasonable pseudo-tPDF data, we can develop the program of Sec.~\ref{sec:Formalism} and extract the desired tPDF. We thus follow that strategy and perform a combined fit to all data points using the functional form in Eq.~\eqref{eq:FitAnsatz}. Note that the real part of the pseudo-tPDF is associated to the valence component, which is even in  $\nu$. We thus retain only even powers of Ioffe time. Furthermore, with the aim of handling the continuum limit extrapolation, we split the fit coefficients as
\begin{equation}
\widetilde{A}_{k}(t,z^{2})=\left.\widetilde{A}_{k}(t,z^{2})\right|_{\textrm{Cont.}}+a L_{k}(t,z^{2})=\widetilde{a}_{k}(t,z^{2})+a L_{k}(t,z^{2})
\end{equation}
where $L_{k}$ are coefficient functions casting the lattice spacing dependence of the fit coefficients $\widetilde{A}_{k}$.

The resulting fit is shown in the left panel of Fig.~\ref{fig:results}. When five powers of $\nu$ are retained, we obtain a $\chi^{2}/\textrm{dof}=95.29/114$. If the quality of the fit remains satisfactory, we find the two higher-order coefficients ($k=6,8$) to be compatible with zero. This observation can be understood by noting that, in practice, the hierarchy of coefficients in the series expansion probes regions of increasing Ioffe time; and that data from the ensemble $\textrm{N6}$, which introduces a third lattice spacing, reaches up to $\nu\simeq 3.5$. As a consequence, the continuum extrapolation of the coefficients sensible to the region of larger Ioffe time becomes unreliable. This is also the reason why the continuum limit results shown in the right panel of Fig.~\ref{fig:results} are restricted to a region $\nu\leq 3$.

\section{Summary and conclusions}

In this proceeding we compute the $\Peta_{c}$-meson tPDF on a lattice QCD setup. We employ the pseudo-distribution approach and propose the description of both, pseudo and light-cone distributions, as power series in Ioffe time. This idea is model-independent and arises from the formal definition of these distributions. We illustrate the procedure using a set of \texttt{CLS} of ensembles, describing the path towards the generation of pseudo-tPDF data and extracting the companion light-cone distribution. We present results for the valence-quark contribution to the Ioffe-time distribution at a renormalization scale of $3~\textrm{GeV}$.

\section*{Acknowledgments}

The authors are grateful to F. de Soto, H. Dutrieux, M. Fucilla, M. Magin-Brinet, S. Mukherjee, J. Rodr\'iguez-Quintero, L. Szymanowski, S. Wallon and S. Zafeiropoulos for valuable comments and discussions. This work has been supported by P2IO LabEx (ANR10-LABX-0038) in the the framework of Investissements d'Avenir (ANR-11-IDEX-0003-01) (JMMC); and by Agence Nationale de la Recherche through contract ANR-17-CE31-0019 (TSJ). Similarly, the authors acknowledge RES for the access provided to HPC resources by BSC in MareNostrum5 to activity FI-2024-2-0014; and also CCRT-TGCC (2023-A0100502271) by GENCI and DRF.

\bibliographystyle{unsrt}
\bibliography{bibliography.bib}

\begin{thebibliography}{10}

\bibitem{Diehl:2003ny}
M.~Diehl.
\newblock {Generalized parton distributions}.
\newblock {\em Phys. Rept.}, 388:41--277, 2003.

\bibitem{Belitsky:2005qn}
A.~V. Belitsky and A.~V. Radyushkin.
\newblock {Unraveling hadron structure with generalized parton distributions}.
\newblock {\em Phys. Rept.}, 418:1--387, 2005.

\bibitem{Burkardt:2000za}
M.~Burkardt.
\newblock {Impact parameter dependent parton distributions and off forward
  parton distributions for $\zeta\to 0$}.
\newblock {\em Phys. Rev.}, D62:071503, 2000.
\newblock [Erratum: Phys. Rev.D66,119903(2002)].

\bibitem{Cichy:2021ewm}
K.~Cichy.
\newblock {Overview of lattice calculations of the x-dependence of PDFs, GPDs
  and TMDs}.
\newblock {\em EPJ Web Conf.}, 258:01005, 2022.

\bibitem{Radyushkin:2017cyf}
A.~V. Radyushkin.
\newblock {Quasi-parton distribution functions, momentum distributions, and
  pseudo-parton distribution functions}.
\newblock {\em Phys. Rev. D}, 96(3):034025, 2017.

\bibitem{Ishikawa:2017faj}
T.~Ishikawa, Y.-Q. Ma, J.-W. Qiu, and S.~Yoshida.
\newblock {Renormalizability of quasiparton distribution functions}.
\newblock {\em Phys. Rev. D}, 96(9):094019, 2017.

\bibitem{Braun:1994jq}
V.~Braun, P.~Gornicki, and L.~Mankiewicz.
\newblock {Ioffe - time distributions instead of parton momentum distributions
  in description of deep inelastic scattering}.
\newblock {\em Phys. Rev. D}, 51:6036--6051, 1995.

\bibitem{Bhattacharya:2023ays}
S.~Bhattacharya, K.~Cichy, M.~Constantinou, X.~Gao, A.~Metz, J.~Miller,
  S.~Mukherjee, P.~Petreczky, F.~Steffens, and Y.~Zhao.
\newblock {Moments of proton GPDs from the OPE of nonlocal quark bilinears up
  to NNLO}.
\newblock {\em Phys. Rev. D}, 108(1):014507, 2023.

\bibitem{Radyushkin:1983wh}
A.~V. Radyushkin.
\newblock {On Spectral Properties of Parton Correlation Functions and
  Multiparton Wave Functions}.
\newblock {\em Phys. Lett. B}, 131:179--182, 1983.

\bibitem{Izubuchi:2018srq}
T.~Izubuchi, X.~Ji, L.~Jin, I.~W. Stewart, and Y.~Zhao.
\newblock {Factorization Theorem Relating Euclidean and Light-Cone Parton
  Distributions}.
\newblock {\em Phys. Rev. D}, 98(5):056004, 2018.

\bibitem{Strichartz:1994mt}
R.~Strichartz.
\newblock {\em {A Guide to Distribution Theory and Fourier Transforms}}.
\newblock CRC Press, 1994.

\bibitem{Zhang:2024djl}
H.-C. Zhang and X.~Ji.
\newblock {On convergence properties of GPD expansion through Mellin/conformal
  moments and orthogonal polynomials}.
\newblock 8 2024.

\bibitem{Fritzsch:2012wq}
P.~Fritzsch, F.~Knechtli, B.~Leder, M.~Marinkovic, S.~Schaefer, R.~Sommer, and
  F.~Virotta.
\newblock {The strange quark mass and Lambda parameter of two flavor QCD}.
\newblock {\em Nucl. Phys. B}, 865:397--429, 2012.

\bibitem{DellaMorte:2017dyu}
M.~Della~Morte, A.~Francis, V.~G\"ulpers, G.~Herdo\'\i{}za, G.~von Hippel,
  H.~Horch, B.~J\"ager, H.~B. Meyer, A.~Nyffeler, and H.~Wittig.
\newblock {The hadronic vacuum polarization contribution to the muon $g-2$ from
  lattice QCD}.
\newblock {\em JHEP}, 10:020, 2017.

\bibitem{Balasubramamian:2019wgx}
R.~Balasubramamian and B.~Blossier.
\newblock {Decay constant of $B_s$ and $B^*_s$ mesons from $\mathrm{N_f}=2$
  lattice QCD}.
\newblock {\em Eur. Phys. J. C}, 80(5):412, 2020.

\bibitem{Jay:2020jkz}
W.~I. Jay and E.~T. Neil.
\newblock {Bayesian model averaging for analysis of lattice field theory
  results}.
\newblock {\em Phys. Rev. D}, 103:114502, 2021.

\end{thebibliography}

\end{document}